# Analysis of the Impact of High-Frequency Trading on Artificial Market Liquidity

Isao Yagi, *Member, IEEE*, Yuji Masuda, and Takanobu Mizuta

*Abstract*—Many empirical studies have discussed market liquidity, which is regarded as a measure of a booming financial market. Further, various indicators for objectively evaluating market liquidity have also been proposed and their merits have been discussed. In recent years, the impact of high-frequency traders (HFTs) on financial markets has been a focal concern, but no studies have systematically discussed their relationship with major market liquidity indicators, including volume, tightness, resiliency, and depth. In this study, we used agent-based simulations to compare the major liquidity indicators in an artificial market where an HFT participated was compared to one where no HFT participated. The results showed that all liquidity indicators in the market where an HFT participated improved more than those in the market where no HFT participated. Furthermore, as a result of investigating the correlations between the major liquidity indicators in our simulations and the extant empirical literature, we found that market liquidity can be measured not only by the major liquidity indicators but also by execution rate. Therefore, it is suggested that it could be appropriate to employ execution rate as a novel liquidity indicator in future studies.

*Index Terms*—Agent-based simulation, artificial market, financial market, market liquidity, multiagent system (MAS).

## I. Introduction

PRESENTLY, market liquidity is considered to be an important factor by numerous investors. Typically, a market is said to be significantly liquid if participants are able to rapidly buy or sell the desired amounts at costs similar to the current market price [1].

Empirical studies on market liquidity have produced various beneficial findings [2]–[27]. Different means of assessing the liquidity of a market were proposed in [2], based on depth, resiliency, and tightness. Many of the other empirical studies, for example, [3]–[7], among others, concern dimensions of market liquidity, that is depth, tightness, and resiliency. In addition to these dimensions, an approach to determining illiquidity (ILLIQ) was also established in [8] based on a ratio of the absolute value of the daily return of a stock to the daily dollar volume, with averaging over a specific time span. Amihud's work additionally demonstrated that illiquidity is associated with increased expected asset returns.

Muranaga [9] performed a cross-sectional assessment of the Tokyo stock exchange (TSE) with the aim of determining the correlations between three different market liquidity parameters (depth, tightness, and resiliency) and trading frequency. In this prior work, market impact, the spread between bid and ask prices, and the rate of convergence of this spread following trades were respectively used as measurements of depth, tightness, and resiliency. The results of this prior analysis established that trade frequency exhibited correlations with these parameters. Muranaga [9] additionally examined the effect of lower TSE tick sizes on liquidity, and reported that a reduced tick size was associated with lower spreads, greater trading frequencies, and decreased volatility. These results confirmed that tick size can have an effect on market liquidity. The correlation between tick size and the quality of the market was established by Chung *et al.* [10] based on data obtained from a market having varying tick sizes. Their study examined the effect of small variations in price on liquidity by assessing inter-temporal and cross-sectional correlations between market liquidity and tick size. Chung's work also showed that there are positive correlations between depth and both turnover rate and volume, but inverse relationships between depth and both volatility and market price. Harris [11] noted that tick size reduction would reduce quoted spreads while causing a reduction in depth and additionally Bessembinder [12] noted that the quoted spread and effective spread would decline because of tick size reduction, as well as market volatility. Ahn *et al.* [13] reported that the quoted spread and effective spread decreased due to the change in tick size implemented by the TSE in 1998. Olbrys and Mursztyn [3] investigated depth, tightness, and resiliency using high-frequency data for 53 companies in the Warsaw stock exchange (WSE). They reported that depth and resiliency for the most liquid big companies with the largest market capitalization were higher than those for the other companies, whereas tightness for the former companies was lower than that for the other companies. Moreover, they found that tightness was high (low) when depth was high (low); however, the majority of the correlation coefficients between tightness and depth were not significantly different from zero, so these indicators can capture various sources of market liquidity.

It should be noted that liquidity is frequently defined differently depending on the research that is being performed. As an example, a scenario in which the sales and purchases in

Manuscript received February 29, 2020; revised July 9, 2020; accepted August 18, 2020. This work was supported by JSPS KAKENHI under Grant 20K04977. *(Corresponding author: Isao Yagi.)*

Isao Yagi is with the Department of Information and Computer Sciences, Faculty of Information Technology, Kanagawa Institute of Technology, Atsugi 243-0292, Japan (e-mail: iyagi2005@gmail.com).

Yuji Masuda is with the System Development, Information System Group, Card Solution Company, Yamato System Development Company Ltd., Kawasaki 211-0004, Japan.

Takanobu Mizuta is with the Investment and Research Division, SPARX Asset Management Company Ltd., Minato 108-0075, Japan.

Digital Object Identifier 10.1109/TCSS.2020.3019352

a market have minimal effect on price and one in which the volatility is low can both be considered to represent markets with high liquidity. Consequently, various studies use different liquidity indicators and so it can be challenging to directly compare the results of different research projects.

Due to IT innovations, the transactions of high-frequency traders (HFTs) now have a great impact on the market. HFTs use powerful computer programs to transact a large number of orders in fractions of a second. As they place both a buy and a sell order to make a profit—by which the sell price exceeds the buy price for an asset—they also play a market maker role. Various HFT strategies have been reported [28], [29]. ASIC [28] categorized these strategies into three categories: 1) electronic liquidity provision strategies; 2) statistical arbitrage strategies; and 3) liquidity detection strategies. A market making strategy is a kind of electronic liquidity provision strategy and majority of the HFT trading volume and more than 80% of HFT limit order submissions are associated with market making strategies [30]. Recently, the rate of transactions attributable to HFTs compared to the total number of orders in financial markets has become large. It is estimated that 48.4% of orders in the foreign exchange market between the U.S. dollar and Japanese yen are HFT orders [31]. Empirical evidence has been put forward to suggest that the transactions of HFTs have increased depth [29]. However, there are some critical opinions such as that HFTs were responsible for the flash crash in the US stock market in 2010.

Thus, there are some unclear points about the effect of HFTs on market liquidity. The reason why is that the empirical data contain many external factors, so that even if some kind of impact can be confirmed in the data, we cannot confirm whether HFT transactions were the cause.

One method for dealing with situations that defy analysis by previous empirical research methods is to construct an artificial market using an agent-based model [32]–[36]. In recent years, much research using multi-agent systems has been performed and achieved many results not only in the financial field but also in other fields, such as cognitive architecture [37], population dynamics [38], epidemiology [39], and social networks [40], to name a few. In this model, specific agents are given unique trading assignments (such as only selling or buying) and subsequently act as investors to perform trading of financial assets. The behavior of these agents in the market is subsequently monitored. Using this technique in conjunction with various market restrictions (such as applying limitations that promote efficiency and stability, including regulations regarding short selling), the behavior of investors can be assessed together with the effects of investor actions on the market [41]–[43].

Some artificial market simulations have been fruitful in terms of producing useful findings [44], [45]. For example, Yagi *et al.* [46] set out to investigate the relationship among liquidity indicators by changing artificial market factors. Specifically, they explored the relationship among four typical indicators—volume, tightness, resiliency, and depth—by changing artificial market factors such as the tick size of market price.

There has already been some discussion on the use of an artificial market that has implemented HFTs [47]–[49]. There have also been many empirical studies that discussed relationships between high-frequency trading and stock market liquidity indicators [29], [30]. However, to the best of the authors' knowledge, no study has hitherto explored the relationships between liquidity indicators in an artificial market context which has implemented HFTs.

In this study, we investigated changes in well-known market liquidity indicators and their correlations in markets where an HFT participated and compared the results with those in markets where an HFT did not participate as shown by Yagi *et al.* [46]. We found that, when the market was stable, HFT transactions had advantageous effects on all four market liquidity indicators: volume, tightness, resiliency, and depth. The results mean that market liquidity is supplied by HFT transactions. Furthermore, we also found that the reason why volume correlated with depth in actual markets is not only that the variation in the order prices of investors changed dynamically [46], but also that the order frequency of investors changed dynamically. This finding indicates that market liquidity can be measured not only by major market indicators such as volume and depth, but also by the execution rate. Therefore, we observe that execution rate may be appropriate as a new market liquidity indicator.

It is difficult to investigate changes in market liquidity indicators and their correlations between the market where an HFT participates and the market where an HFT does not participate using empirical data. This is because, as already mentioned, the empirical data contain many external factors and it is difficult to identify the HFT transactions among the various trader transactions. Moreover, even if HFT transactions can be identified and removed from the original empirical data, the remaining data cannot be regarded as data of the market where HFTs do not participate. This is because the investment behavior of other investors is likely to be affected by HFT transactions. Actual data are available when HFTs did not exist yet; however, it is difficult to compare the current market data with HFTs with the market data at that time, as the composition of external factors in the current market is different from that in the market at that time. Our study and Yagi *et al.* [46] confirmed the difference of the change of market liquidity indicators among markets where the trading strategies, which are fundamental, technical, and noise, that investors focus on are different. To compare the simulation results with actual data and investigate whether the results are valid, it is necessary to measure the number of investors using these strategies from actual data. However, that is potentially an enormous task. On the other hand, an artificial market can deal with some issues that empirical study methods are unsuitable for, as it is possible to investigate the pure effect of a specific factor on the market by changing only the specific factor and fixing all other factors in the artificial market.

As has been noted, the validation of artificial market models with actual data requires much work, and it would be a separate research project [35], [50], [51]. LeBaron [50] and Chen *et al.* [35] investigated the validation of previous artificial market models by changing various factors in the models.

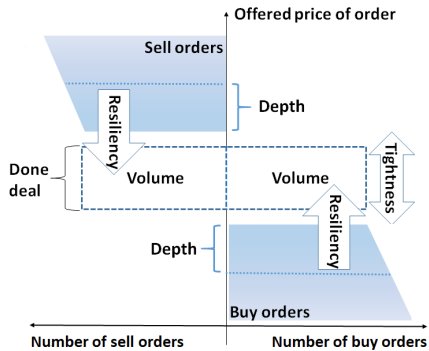

Fig. 1. Indicators of market liquidity in view of buy and sell orders.

Hirano *et al.* [51] proposed an HFT model for an artificial market, compared the simulation results obtained from their model with data that seemed to be HFT transactions exacted from actual data, and found some inconsistencies between the two. Solving this problem will take more time because it is necessary to check whether the extracted data are really only HFT transactions, whether the HFT model is inappropriate, etc. Therefore, in this study, we focused on investigating how the HFT orders affect market liquidity and the relationship between the market liquidity indicators by using a simulation model that is simple but captures the characteristics of the market and the traders.

In what follows, first, we explain the definition of market liquidity in Section II. Next, we explain the proposed artificial market used in the present study in Section III. Then, in Section IV-C1, we investigate changes in four well-known market liquidity indicators and their correlations in markets where an HFT participated, and compare the results to markets where no HFT participated as shown by Yagi *et al.* [46]. Hereafter, we term the market where an HFT participates "the market with HFT" and the market where an HFT does not participate "the market without HFT." Furthermore, we discuss the reason why execution rate may be appropriate as a new market liquidity indicator based on the simulation results in which the order frequency of agents is changed in Section IV-C2. Finally, we offer conclusions including suggestions for future work in Section V.

## II. MARKET LIQUIDITY AND LIQUIDITY INDICATORS

There are different definitions and understandings of market liquidity in the extant literature. Even so, liquidity is typically defined as a scenario in which buyers and sellers can quickly perform transactions involving significant amounts without overly affecting price [15]. Prior research performed on an empirical basis [16] employed tightness, volume, depth, and resiliency to assess market liquidity (see Fig. 1).

### A. Volume

Volume is an indicator of liquidity that reflects the quantity of trading in a market. Trading volume is traditionally used to measure the existence of numerous market participants and transaction [16], [26]. This factor can be quantified using a parameter, turnover, defined as the quantity of shares that are traded throughout the overall market over a specific time span. This is possible because orders that are both numerous and frequent, and that occur over a brief period, are more readily matched in conjunction with a high turnover. Thus, liquidity will exhibit a positive correlation with turnover/volume. Nishizaki *et al.* [16] and Sarr and Lybek [26] employed turnover and turnover rate to measure market liquidity. In the work reported herein, turnover, that is Volume, was defined as the volume of trading in each run of the simulation, as follows:

$$\text{Volume} = \sum_{t=0}^{t_{\text{end}}} V^t \quad (1)$$

where $t_{\text{end}}$ is the end time of the simulation and $V^t$ is trading volume at time $t$.

### B. Tightness

Tightness is defined as the extent to which the costs of transactions are minimized, including the gap between selling and buying prices. Tightness can be quantified as the spread between the asking and bidding prices. That is, the difference between the best-ask (the minimum sell order pricing that appears in the order book) and the best-bid (the maximum buy order pricing) for any given market asset. Based on this definition, higher liquidity will be associated with lower bid-ask spreads. Consequently, Tightness can be determined by taking the mean of the bid-ask spreads generated during runs of the simulation, as follows:

$$\text{Tightness} = \frac{1}{t_{\text{end}}} \sum_{t=0}^{t_{\text{end}}} \left( P_{ba}^t - P_{bb}^t \right) \quad (2)$$

where $P_{ba}^t$ and $P_{bb}^t$ are the best-ask price and best-bid price at time $t$, respectively.

It is noted that the empirical literature contains examples of different kinds of bid-ask spreads for measuring tightness. Von Wyss [17] and Olbrys and Mursztyn [3] employed relative spread as a measure of tightness. Goyenko *et al.* [4] and Fong *et al.* [5] used the effective spread and the realized spread on the basis of Huang and Stoll [27] to measure liquidity.

### C. Resiliency

Resiliency reflects the ability of a market to permit the rapid flow of new orders such that imbalances are corrected. This prevents the shifting of prices away from the values expected from fundamental principles.

Resiliency can be quantified as the ratio of the price range to the turnover on a daily basis [16], and is calculated as the gap between the highest and lowest costs of transactions in a given day over the turnover in a trading day. A more highly resilient market is typically associated with lower values of this ratio, which can also assist traders in performing efficient transactions. Herein, Resiliency is defined as the ratio of the mean price range to the turnover each day [16] throughout each run of the simulation, subsequently denoted as the P/T ratio. Thus, higher liquidity is reflected in a reduction in the P/T ratio.

Besides the P/T ratio, there are various versions of measures for resiliency in the literature. Olbrys and Mursztyn [3] and Goyenko *et al.* [4] employed realized spread as a measure of resiliency. It is noted that defining an indicator to measure resiliency is difficult [18]. Empirical studies have proposed various methodologies for Resiliency. For example, some studies employed methodologies based on adopted VAR [19]–[21] and the Kalman filter with the ARMA approach [22]. Olbrys and Mursztyn [6] employed a new methodology for stock market resiliency measurement based on the discrete Fourier transform for high-frequency intraday data. Moreover, they also proposed a methodology for stock market resiliency measurement based on the short-time Fourier transform for high-frequency intraday data [7].

In this article, Resiliency was determined by the P/T ratio for simplicity. The P/T ratio is calculated as follows:

$$P/T \text{ratio} = \frac{1}{t_{\text{end}}/t_{\text{day}}} \sum_{t^d=1}^{t_{\text{end}}/t_{\text{day}}} \frac{P_h^{t^d} - P_l^{t^d}}{V^d} \quad (3)$$

where $t_{\text{day}}$ is number of time steps in one day of the simulation and $P_h^{t^d}$ and $P_l^{t^d}$ are the highest and lowest prices of the $t^d$th day, respectively.

### D. Depth

The quantity of orders in the highest buy order price range and the lowest sell order price range can be used as a definition of Depth [17], [18], [23], [24]. This parameter is thus related to the quantity of orders from potential sellers or buyers that are either readily uncovered or actually occur and that are either below or above the current trading price. Based on this definition, a greater Depth will correlate with a higher degree of liquidity. In previous empirical studies, various proxies were employed as depth: for example, order ratio [3], [18], total number of limit orders posted at the bid and ask prices [23], average depth of the buy orders and the sell orders in the order book [17], [25], and dollar depth calculated in currency terms [17], [25].

In the work reported herein, data regarding Depth were obtained from the trading board, based on the total number of sell orders in the range between the lowest price and 50 ticks above this price and the total number of sell orders between the highest price and 50 ticks below this price. Here, a tick represents the smallest pricing unit (refer to Section III). Then depth is given by the following equation:

$$\text{Depth} = \sum_{t=0}^{t_{\text{end}}} \sum_{i=0}^{n} \frac{B\text{Order}_{bb-i\cdot\Delta P}^t + S\text{Order}_{ba+i\cdot\Delta P}^t}{2} \quad (4)$$

where $B\text{Order}_{bb-i\cdot\Delta P}^t (S\text{Order}_{ba+i\cdot\Delta P}^t)$ is the number of buy(sell) orders whose price are at the best bid(ask) price plus(minus) $i \cdot \Delta P$, i.e., $\Delta P$ is the tick size, of in the order book and $n$ is 49.

## III. ARTIFICIAL MARKET MODEL

We have built a new artificial market model with an HFT agent on the basis of the artificial market model put forward by Yagi *et al.* [46], because their model can reproduce the statistical characteristics of the kinds of long-term price fluctuations observed in empirical analyses. We seek to investigate the relationship among the four market liquidity indicators by changing the parameters. For simplicity, the HFT agent is modeled on the basis of Kusada *et al.* [52], which is a market making strategy. The model constructed in the present work has a single risk asset associated with trading, which is simply referred to as an asset, and a non-risk asset is referred to as cash.

The model employs a continuous trading session (that is, a double auction), in which prices are determined based on the presence of buy or sell order prices within the order book that are above or below the agent's sell or buy target. In such cases, the order from the agent is instantly assigned to the highest buy or lowest sell order. This process is designated as a market order. In the case that the order book does not contain any orders matching these descriptions, the orders remain in the book. This scenario is designated as a limit order. These orders will be negated at a specific time, $t_c$, (termed the order effective period) following placement of the order. The minimum pricing unit, $\Delta P$, equals the tick size, and when orders are sell orders, fractional values smaller than $\Delta P$ are rounded up. By contrast, when orders are buy orders, they are rounded down.

There are two types of agents in our model, namely a normal agent as a general investor and an agent as an HFT. It is noted that an agent does not correspond to an actual investor, but represents a similar trading strategic group. First, we explain the normal agent model. In the case that the quantity of normal agents is $n$, each agent $j = 1, \ldots, n$ submits an order in turn. Following the order from agent $n$, agent 1 submits another order. With each order, the time value, $t$, is increased by 1 which represents a stepwise progression even in the case that there is no trade, such that the order is transferred to the order book. In this mechanism, a single order is placed by each agent in each step and these agents have essentially infinite cash assets. These agents are also permitted to short sell.

The price of an order by agent $j$ in a given transaction is calculated using the following process. Firstly, the rate at which the price anticipated by this agent changes at a given time, $t$, (equivalent to the expected return) is designated $r_{e_j}^t$ and calculated using the equation

$$r_{e_j}^t = \frac{1}{w_{1,j}^t + w_{2,j}^t + u_j}\left(w_{1,j}^t r_{e_{1,j}}^t + w_{2,j}^t r_{e_{2,j}}^t + u_j \epsilon_j^t\right) \quad (5)$$

where $w_{i,j}^t$ equals the weight of the $i$th term associated with agent $j$ at a given time $t$, the value of which is determined based on a uniform distribution between 0 and $w_{i,\max}$ at the simulation onset and subsequently modified via the learning mechanism discussed below. In addition, $u_j$ equals the third term weight and is based on a uniform distribution between 0 and $u_{\max}$ at the simulation onset but remains unchanged. The degree of each trading strategy (that is, technical, fundamental, and noise) is assumed to disperse across all agents, as detailed in the following discussion. Consequently, the trading weights were modeled as independently selected random variables.

In (5), the first term on the right is used to normalize the effects of the various trading strategies. The initial term in the brackets on the right-hand side of this equation is associated with the fundamental strategy, such that a negative (positive) return is anticipated by the agent in the case that the market price is higher (lower) than the fundamental price. Here, $r_{e_1,j}^t$ represents the return expected when using the fundamental strategy in the case of agent $j$ at a given time $t$. This value is equivalent to $\ln(P_f/P^{t-n})$, where $P_f$ equals the fundamental price (that is unchanged over time) while $P^t$ is the market price at a given time $t$. In the absence of trading, the latter value is set equal to the price most recently quoted. The initial market price is given the same value as the fundamental price, meaning that $P^0 = P_f$.

The central term is associated with the technical strategy, for which the agent anticipates a negative (positive) return in the case that the historical return is also negative (positive). The term $r_{e_2,j}^t$ is the anticipated return resulting from this strategy at time $t$ for the agent, and equals $\ln(P^{t-n}/P^{t-n-\tau_j})$. Here, $\tau_j$ is given a value based on a uniform distribution ranging from 1 to $\tau_{\max}$ at the simulation onset. It seems natural to think that investors calculate their historical returns based on prices at their preferred time. That is, we think that each investor uses different historical returns. Thus, we employed $\tau_j$ to model the historical return for each agent.

The final (third) term is associated with the noise strategy. In this term, $\epsilon_j^t$ equals the random error, which has a Gaussian distribution with a mean of zero and a standard deviation of $\sigma_\epsilon$.

It is noted that $r_{e_j}^t$ does not predict the return of agent j at time $t+1$, but predicts the return that can be expected in the future. In other words, agents determine the order price not considering when the return that they expect will be achieved. Instead, they determine the order price based on the idea that the return that they expect will be achieved eventually.

Based on the expected return $r_{e_j}^t$, the expected price $P_{e_j}^t$ is found using the following equation:

$$P_{e_j}^t = P^{t-1} \exp(r_{e_j}^t). \tag{6}$$

The order price $P_{o_j}^t$ is a normally distributed random number with mean $P_{e_j}^t$ and standard deviation $P_\sigma^t$ given by

$$P_\sigma^t = P_{e_j}^t \cdot \text{Est} \tag{7}$$

where, for expedience, Est($0 < \text{Est} \leq 1$) refers to the variation coefficient of the order price[1]. The choice between buying and selling is determined by the relative sizes of the expected price $P_{e_j}^t$ and the order price $P_{o_j}^t$. That is, if $P_{e_j}^t > P_{o_j}^t$, an agent places a buy order for one share, else if $P_{e_j}^t < P_{o_j}^t$, an agent places a sell order for one share. Finally, if $P_{e_j}^t = P_{o_j}^t$, an agent places no order.

Previous studies using artificial markets have implemented various kinds of learning processes. For example, agents switch strategies and/or tune their strategy parameters based on their performance, market price, and so on [33], [53], [54].

[1] Est was prepared to model a market composed of ordinary investors with high or low confidence in the expected price. In other words, investors who are confident in their expected prices may place their orders near the expected price, but investors who are uncertain about their expected price, for example, due to the uncertain external environment, may place their orders at prices that deviate from the expected price. However, HFTs employ a market making strategy that is a mechanically ordering strategy based on a rule. Therefore, as we shall see later, the HFT agent is not affected by Est.

The learning process in the present study is implemented to switch between the fundamental and technical strategies.

We modeled the learning process as follows based on Yagi et al. [44]. For $r_{e_i,j}^t$, learning occurs by each agent immediately before the agent places an order. That is, when $r_{e_i,j}^t$ and $r_l^t = \ln(P^{t-1}/P^{t-t_l})$ are of the same sign, $w_{i,j}$ is updated as follows:

$$w_{i,j}^t \leftarrow w_{i,j}^t + k_l|r_l^t|q_j^t(w_{i,\max} - w_{i,j}^t) \tag{8}$$

where $k_l$ is a constant, and $q_j^t$ is set according to the uniform distribution between 0 and 1. When $r_{e_i,j}^t$ and $r_l^t$ have opposite signs, $w_{i,j}$ is updated as follows:

$$w_{i,j}^t \leftarrow w_{i,j}^t - k_l|r_l^t|q_j^t w_{i,j}^t. \tag{9}$$

Separately from the process of learning based on past performance, $w_{i,j}^t$ is reset with a small probability $m$, according to the uniform distribution between 0 and $w_{i,\max}$.

Next, we will explain the HFT agent model in detail. Among the market making strategy models that have been proposed [51], [52], [55], we modeled the HFT agent on the basis of Kusada et al. [52] for simplicity. There is only one HFT agent. The HFT agent places both a buy limit order and a sell limit order as each normal agent places its order. If the previous sell or buy or both orders of the HFT agent remain in the order book, the HFT agent cancels them and places new buy and sell limit orders. Generally, an HFT agent decides its own order price based on the best-bid, the best-ask, and the spread which is equal to the amount of its own expected return per transaction. However, the order price of the HFT agent also depends on its position, which means the amount of an asset held by the HFT agent, as it acts to keep its position neutral. That is, when the HFT agent has a long position, which means the agent buys and holds some amount of an asset, its buy and sell order prices are set lower so that its sell order matches an order from normal agents easier than its buy order. On the other hand, when the HFT agent has a short position, which means the agent short-sells the asset, its buy and sell order prices are set higher so that its buy order matches an order from normal agents easier than its sell order [55], [56]. Let the base spread of the HFT agent and the coefficient of its position (its initial value is set based on Kusaka et al. [55] as $5.0 \times 10^{-8}$) be $\theta_h$ and $w_h$. Let the best-bid, the best-ask, its basic order price, its buy order price, its sell order price at time $t$, and the HFT's position between time $t$ and $t+1$ be $P_{bb}^t$, $P_{ba}^t$, $P_{bv,h}^t$, $P_{bo,h}^t$, $P_{so,h}^t$, and $s_h^t$, respectively. Then, $P_{bo,h}^t$, $P_{so,h}^t$, and $P_{bv,h}^t$ are as follows:

$$P_{bo,h}^t = P_{bv}{}_H^t - \frac{1}{2}P_f \cdot \theta_h \tag{10}$$

$$P_{so,h}^t = P_{bv}{}_H^t + \frac{1}{2}P_f \cdot \theta_h \tag{11}$$

$$P_{bv,h}^t = (1 - w_h(s_h^t)^3) \cdot \frac{1}{2}(P_{bb}^t + P_{ba}^t). \tag{12}$$

When the sell (buy) order price of the HFT agent is lower (higher) than the best-bid (best-ask), the HFT agent's order becomes a market order. Therefore, if the following conditions are satisfied, the buy and sell order prices of the

TABLE I
PARAMETERS

| Parameter | Initial value |
|---|---|
| $t_{end}$ | 1,000,000 |
| $n$ | 1,000 |
| $w_{1,max}$ | 1 |
| $w_{2,max}$ | 10 |
| $u_{max}$ | 1 |
| $\tau_{max}$ | 10,000 |
| $\sigma_\epsilon$ | 0.06 |
| $Est$ | 0.003 |
| $t_c$ | 20,000 |
| $\Delta P$ | 0.1 |
| $P_f$ | 10,000 |
| $k_l$ | 4 |
| $m$ | 0.01 |
| $\theta_h$ | 0.002 |
| $w_h$ | $5.0 \times 10^{-8}$ |
| $\Pr_o$ | 1.0 |

TABLE II
PARAMETERS AND SETTINGS

| Parameter | Values | | | | |
|---|---|---|---|---|---|
| $\Delta P$ | 0.01 | 0.1 | 1.0 | 10 | 100 |
| $w_{1,max}$ | 1.0 | 3.0 | 5.0 | 8.0 | 10.0 |
| $w_{2,max}$ | 1.0 | 3.0 | 5.0 | 8.0 | 10.0 |
| $\sigma_\epsilon$ | 0.02 | 0.04 | 0.06 | 0.08 | 0.1 |
| $Est$ | 0.003 | 0.005 | 0.01 | 0.02 | 0.03 |
| $t_c$ | 10,000 | 15,000 | 20,000 | 25,000 | 30,000 |
| $\Pr_o$ | 0.2 | 0.4 | 0.6 | 0.8 | 1.0 |

HFT agent are changed [52]. That is, if $P_{bo,h}^t \geq P_{ba}^t$, then

$$P_{bo,h}^t = P_{ba}^t - \Delta P \quad (13)$$
$$P_{so,h}^t = \left(P_{ba}^t - \Delta P\right) + P_f \cdot \theta_h. \quad (14)$$

If $P_{so,h}^t \leq P_{bb}^t$, then

$$P_{bo,h}^t = \left(P_{bb}^t + \Delta P\right) - P_f \cdot \theta_h \quad (15)$$
$$P_{so,h}^t = P_{bb}^t + \Delta P. \quad (16)$$

## IV. SIMULATION RESULTS AND DISCUSSION

### A. Overview

We show how the four liquidity indicators, execution rate, and volatility in both the market with HFT and the market without HFT change when one parameter is changed. We set the initial values of the model parameters as listed in Table I. In the simulations, various parameters are then changed as listed in Table II.[2] For each simulation, only one parameter is changed and the others are maintained at their initial values. We ran simulations from $t = 0$ to $t = t_{end}$, and the four liquidity indicator values, execution rate, and volatility were calculated as averages over 100 simulation runs. We define $t_{day} = 20\,000$ time steps as 1 day because the number of trades for $20\,000$ time steps is approximately the same as that in actual markets in 1 day.

---
[2]We built our artificial market to be simple, but it reproduced the characteristics of general financial markets, that is the stylized facts. Therefore, we can say that most of the parameters of our artificial market, namely the parameters in Table II, are important. Thus, we varied all of these parameters one at a time in order to investigate the four liquidity indicators, execution rate, and volatility in both the market with HFT and the market without HFT.

TABLE III
STYLIZED FACTS

| Kurtosis | | 0.802 |
|---|---|---|
| | Lag | |
| Autocorrelation | 1 | 0.206 |
| coefficients | 2 | 0.132 |
| for squared returns | 3 | 0.103 |
| | 4 | 0.088 |
| | 5 | 0.078 |

In Section IV-B, we validate our basic artificial market model whose parameters are set as listed in Table I. In Section IV-C1, we investigate changes in the four market liquidity indicators and their correlations between the market with HFT and the market without HFT. In Section IV-C2, we discuss the reason why execution rate appears to be an appropriate new market liquidity indicator based on the simulation results in which the order frequency of agents is changed.

### B. Validation of Proposed Artificial Market

As many empirical studies have noted, particular statistical properties such as a fat tail and volatility clustering appear in actual markets [57]–[59]. They showed that kurtosis of returns and autocorrelation coefficients for squared returns with several lags are positive when a fat tail and volatility clustering appear in actual markets. For example, the kurtoses of monthly log returns of U.S. bonds, the S&P composite index of 500 stocks, and Microsoft are 4.86, 7.77, and 1.19, respectively [59]. It is noted that kurtosis of returns depends on the kinds of financial assets. Likewise, autocorrelation coefficients for squared returns with several lags of the S&P composite index of 500 stocks are 0.0536, 0.0537, 0.0537, 0.0538, and 0.0538 when lags are from 1 to 5, respectively [59].

One of the objectives of artificial markets is to reproduce the statistical features of the price process with minimal hypotheses about the intelligence of agents [60]. Early artificial markets appeared unable to account for the ubiquitous scaling laws of returns; however, recent artificial markets have been able to reproduce fat tails and volatility clustering [46], [52], [61]–[64]. Therefore, we set the artificial market parameters as listed in Table I as well as Yagi *et al.* [46] so as to replicate these features. The values of these parameters were determined by trial-and-error experiments in Mizuta *et al.* [36], which described the base model used in Yagi *et al.* [46]. Mizuta *et al.* [36] added a learning process to the model of Chiarella *et al.* [34] and reproduced large-scale market confusion such as a bubble or a financial crisis in their artificial market model.

Table III shows the statistics for stylized facts, which are averages over 100 simulation runs, for which we calculated price returns at intervals of 100 time units. Table III shows that both kurtosis and autocorrelation coefficients for squared returns with several lags are positive, which means that all runs replicate a fat tail and volatility clustering. This indicates that the model replicates long-term statistical characteristics observed in real financial markets. Therefore, it can be confirmed that the proposed model is valid.

TABLE IV

LIQUIDITY INDICATORS, EXECUTION RATE, AND VOLATILITY WHEN $\Delta P$ IS CHANGED

|  | with HFT | | | | | | without HFT | | | | | |
| --- | --- | --- | --- | --- | --- | --- | --- | --- | --- | --- | --- | --- |
|  | Volume | Tightness | Resiliency | Depth | Execution rate | Volatility | Volume | Tightness | Resiliency | Depth | Execution rate | Volatility |
| 0.01 | 281,449 (234,518) | 2.938 | 0.00157 | 3,962 | 28.7% | 0.022% | 265,884 | 12.015 | 0.0104 | 2,362 | 27.1% | 0.121% |
| 0.1 | 284,172 (235,545) | 3.042 | 0.00164 | 3,885 | 29.0% | 0.022% | 267,309 | 12.174 | 0.0106 | 2,326 | 27.3% | 0.118% |
| 1 | 277,638 (208,674) | 4.012 | 0.00230 | 3,804 | 28.3% | 0.031% | 265,894 | 12.557 | 0.0106 | 2,369 | 27.1% | 0.121% |
| 10 | 255,839 ( 49,880) | 13.826 | 0.00776 | 3,053 | 26.1% | 0.127% | 266,892 | 17.579 | 0.0118 | 2,431 | 27.2% | 0.148% |
| 100 | 249,194 ( 0) | 100.001 | 0.01933 | 3,183 | 25.4% | 0.705% | 249,419 | 100.001 | 0.019 | 3,189 | 25.5% | 0.708% |

TABLE V

LIQUIDITY INDICATORS, EXECUTION RATE, AND VOLATILITY WHEN $w_{1,\max}$ IS CHANGED

|  | with HFT | | | | | | without HFT | | | | | |
| --- | --- | --- | --- | --- | --- | --- | --- | --- | --- | --- | --- | --- |
|  | Volume | Tightness | Resiliency | Depth | Execution rate | Volatility | Volume | Tightness | Resiliency | Depth | Execution rate | Volatility |
| 1 | 277,638 (208,674) | 4.012 | 0.00230 | 3,804 | 28.3% | 0.031% | 265,894 | 12.557 | 0.0106 | 2,369 | 27.1% | 0.121% |
| 3 | 257,887 (186,728) | 3.947 | 0.00233 | 4,171 | 26.3% | 0.031% | 247,682 | 12.014 | 0.0094 | 2,807 | 25.3% | 0.112% |
| 5 | 239,392 (166,287) | 3.886 | 0.00238 | 4,510 | 24.4% | 0.031% | 230,026 | 11.772 | 0.0084 | 3,283 | 23.5% | 0.110% |
| 8 | 216,706 (141,574) | 3.804 | 0.00248 | 4,926 | 22.1% | 0.031% | 207,453 | 11.171 | 0.0074 | 3,896 | 21.2% | 0.101% |
| 10 | 205,187 (129,045) | 3.763 | 0.00255 | 5,139 | 20.9% | 0.031% | 196,318 | 10.675 | 0.007 | 4,201 | 20.0% | 0.095% |

TABLE VI

LIQUIDITY INDICATORS, EXECUTION RATE, AND VOLATILITY WHEN $w_{2,\max}$ IS CHANGED

|  | with HFT | | | | | | without HFT | | | | | |
| --- | --- | --- | --- | --- | --- | --- | --- | --- | --- | --- | --- | --- |
|  | Volume | Tightness | Resiliency | Depth | Execution rate | Volatility | Volume | Tightness | Resiliency | Depth | Execution rate | Volatility |
| 1 | 406,155 (368,774) | 4.497 | 0.00222 | 1,441 | 41.4% | 0.031% | 372,930 | 78.297 | 0.0518 | 223 | 38.1% | 0.757% |
| 3 | 366,397 (318,034) | 4.330 | 0.00211 | 2,172 | 37.4% | 0.031% | 348,262 | 31.672 | 0.0267 | 645 | 35.5% | 0.299% |
| 5 | 333,144 (276,632) | 4.208 | 0.00212 | 2,783 | 34.0% | 0.031% | 320,592 | 21.473 | 0.018 | 1,080 | 32.7% | 0.204% |
| 8 | 296,810 (231,665) | 4.080 | 0.00224 | 3,453 | 30.3% | 0.031% | 285,224 | 14.869 | 0.0123 | 1,840 | 29.1% | 0.143% |
| 10 | 277,638 (208,674) | 4.012 | 0.00230 | 3,804 | 28.3% | 0.031% | 265,894 | 12.557 | 0.0106 | 2,369 | 27.1% | 0.121% |

TABLE VII

LIQUIDITY INDICATORS, EXECUTION RATE, AND VOLATILITY WHEN $\sigma_\epsilon$ IS CHANGED

|  | with HFT | | | | | | without HFT | | | | | |
| --- | --- | --- | --- | --- | --- | --- | --- | --- | --- | --- | --- | --- |
|  | Volume | Tightness | Resiliency | Depth | Execution rate | Volatility | Volume | Tightness | Resiliency | Depth | Execution rate | Volatility |
| 0.02 | 152,011 ( 67,114) | 3.471 | 0.00335 | 6,116 | 15.5% | 0.030% | 151,109 | 5.308 | 0.0061 | 5,655 | 15.4% | 0.048% |
| 0.04 | 229,627 (152,747) | 3.837 | 0.00254 | 4,693 | 23.4% | 0.031% | 222,479 | 8.975 | 0.0081 | 3,644 | 22.7% | 0.087% |
| 0.06 | 277,638 (208,674) | 4.012 | 0.00230 | 3,804 | 28.3% | 0.031% | 265,894 | 12.557 | 0.0106 | 2,369 | 27.1% | 0.121% |
| 0.08 | 308,131 (245,769) | 4.120 | 0.00218 | 3,239 | 31.4% | 0.031% | 296,523 | 15.928 | 0.0137 | 1,648 | 30.3% | 0.154% |
| 0.1 | 329,797 (271,548) | 4.193 | 0.00216 | 2,832 | 33.7% | 0.031% | 318,035 | 18.821 | 0.0169 | 1,268 | 32.5% | 0.182% |

### C. Results and Discussion

In this section, we investigate the following topics by changing the parameters as shown in Table II:

1) changes in four market liquidity indicators and their correlations in the market with HFT and the market without HFT;
2) difference between the four market liquidity indicators and execution rate.

*1) Comparing Market Liquidity in the Market With HFT and the Market Without HFT:* Tables IV–IX show the four market liquidity indicators (volume, tightness, resiliency, and depth), execution rate of normal agents' orders, and volatility in the market with HFT and the market without HFT by changing the parameters as shown in Table II. Values in parentheses pertain to HFT volume in the total volume and the execution rate means the amount of executed orders in the order book divided by the amount of new orders placed by normal agents during a simulation. In the market with HFT, we can observe that the four market liquidity indicators, and in particular, tightness, resiliency, and depth, are improved to a greater extent than in the market without HFT.

This finding indicates that HFT transactions may improve market liquidity when the asset market is stable in the sense that the fundamental price of the asset is stable.

The reason why the four market liquidity indicators improve is that the bid-ask spread becomes small as the HFT places orders.

The result is directly reflected in tightness. It is noted that tightness is nearly identical in the market with HFT and the market without HFT when the tick size is large. As the minimum of the bid-ask spread is equal to the tick size, the spread between the HFT's buy and sell orders is also forcibly increased when the tick size increases. Therefore, the HFT cannot place orders if the spread is smaller than the bid-ask spread in the market, hence why tightness is similar in the two markets. When the tick size is smaller than the volatility of the market without HFT, which is 0.122% and is about 12 when converted to market price, it does not affect normal agents' behavior. However, when the tick size becomes larger and close to the volatility of the market without HFT, some order prices of normal agents are between market price $P$ and market price $P + \Delta P$. Therefore, if the orders are buy (sell) orders, they are rounded down (up). It is noted that this explanation was also given in Yagi et al. [46].

TABLE VIII
LIQUIDITY INDICATORS, EXECUTION RATE, AND VOLATILITY WHEN Est IS CHANGED

|  | with HFT | | | | | | without HFT | | | | | |
| --- | --- | --- | --- | --- | --- | --- | --- | --- | --- | --- | --- | --- |
|  | Volume | Tightness | Resiliency | Depth | Execution rate | Volatility | Volume | Tightness | Resiliency | Depth | Execution rate | Volatility |
| 0.003 | 277,638 (208,674) | 4.012 | 0.00230 | 3,804 | 28.3% | 0.031% | 265,894 | 12.557 | 0.0106 | 2,369 | 27.1% | 0.121% |
| 0.005 | 220,858 (168,296) | 3.984 | 0.00290 | 3,344 | 22.5% | 0.031% | 206,866 | 12.937 | 0.0125 | 2,405 | 21.1% | 0.122% |
| 0.01 | 146,416 (113,888) | 3.942 | 0.00445 | 2,247 | 14.9% | 0.030% | 131,180 | 13.605 | 0.0191 | 1,821 | 13.4% | 0.127% |
| 0.02 | 84,989 ( 66,955) | 3.898 | 0.00758 | 1,292 | 8.7% | 0.029% | 73,462 | 14.208 | 0.0338 | 1,088 | 7.5% | 0.126% |
| 0.03 | 59,677 ( 47,319) | 3.879 | 0.01074 | 896 | 6.1% | 0.029% | 51,052 | 14.598 | 0.0491 | 754 | 5.2% | 0.126% |

TABLE IX
LIQUIDITY INDICATORS, EXECUTION RATE, AND VOLATILITY WHEN $Pr_o$ IS CHANGED

|  | with HFT | | | | | | without HFT | | | | | |
| --- | --- | --- | --- | --- | --- | --- | --- | --- | --- | --- | --- | --- |
|  | Volume | Tightness | Resiliency | Depth | Execution rate | Volatility | Volume | Tightness | Resiliency | Depth | Execution rate | Volatility |
| 20% | 55,573 ( 42,074) | 3.836 | 0.01120 | 766 | 28.4% | 0.029% | 52,870 | 13.193 | 0.0533 | 492 | 27.0% | 0.118% |
| 40% | 110,591 ( 83,151) | 3.878 | 0.00570 | 1,537 | 28.2% | 0.030% | 105,706 | 12.741 | 0.0265 | 970 | 27.0% | 0.117% |
| 60% | 166,646 (125,485) | 3.924 | 0.00379 | 2,285 | 28.3% | 0.030% | 159,214 | 12.541 | 0.0176 | 1,436 | 27.1% | 0.121% |
| 80% | 222,516 (167,836) | 3.971 | 0.00283 | 3,038 | 28.4% | 0.030% | 212,632 | 12.484 | 0.0133 | 1,893 | 27.1% | 0.119% |
| 100% | 277,638 (208,674) | 4.012 | 0.00230 | 3,804 | 28.3% | 0.031% | 266,461 | 12.360 | 0.0107 | 2,356 | 27.2% | 0.121% |

The reason why Volume increases is that the orders of normal agents are more likely to become market orders; as the bid-ask spread narrows, therefore, the execution rate in the market with HFT increases. It is noted that the increase in volume is limited because the execution rate in the market with HFT does not increase that much compared to the execution rate in the market without HFT.

The reason why resiliency decreases is as follows. As described above, volume, which is the denominator of the P/T ratio, increases slightly, while the difference between the highest and lowest prices of the day, which is the numerator of the P/T ratio, decreases, since the bid-ask spread and volatility decrease. Thus, the P/T ratio decreases.

The increase in depth indicates the amount of HFT orders in the order book that matched orders of normal agents (refer to the values in parentheses in the volume column in Tables IV–VIII). These tables show that most of the executed orders in the order book are HFT orders. As HFT orders in the order book are almost always the best bid or ask orders or both, they usually match the new orders placed by normal agents. On the other hand, the orders of normal agents who did not match the orders in the order book remain in the order book for a long time. Therefore, depth increases.

Next, we explore how market liquidity indicators changes when parameters are changed.

When the tick size increases, only depth in the market with HFT is inversely correlated with depth in the market without HFT (refer to Table IV). The difference between volume in the market with HFT and that in the market without HFT becomes large when the tick size decreases. The smaller the tick size is, the larger the ratio of the amount of the HFT's executed orders in the order book. Thus, most normal agents' orders remain in the order book and depth in the market with HFT increases. On the other hand, when the tick size increases, the HFT's orders do not match any other orders. Therefore, depth in the market with and without HFT do not differ significantly when the tick size is large.

When the maximum fundamental strategy weight, the maximum technical strategy weight, and the standard deviation of the noise factor of the expected return change (refer to Tables V–VII), only resiliency out of the four market liquidity indicators exhibits no positive correlation between the market with and without HFT. The reason is as follows. When these parameters increase, rate of changes in volume both in the market with and without HFT in each table appear to be the same; however, volatility in the market without HFT exhibits a positive correlation with Volume, while volatility in the market with HFT does not correlate with Volume. Therefore, resiliency is more affected by volume in the market with HFT than in the market without HFT by volume; recall that volume is the denominator of the P/T ratio, and the former does not correlate with the latter.

When the variation in order prices changes (refer to Table VIII), tightness in the market with HFT exhibits a negative correlation with Tightness in the market without HFT. The reason is as follows. In the market without HFT, when the variation in order prices increases, the variation in limit order prices increases in the order book and the bid-ask spread also increases [46]. On the other hand, the bid-ask spread in the market with HFT is unlikely to expand, as the spread between the HFT's buy and sell orders is narrow and almost constant. Furthermore, volatility tends to decrease and the bid-ask spread also decreases when the variation in order prices increases.

*2) Execution Rate as Market Liquidity Indicator:* As per the foregoing, it is unlikely in principle that both volume and depth increase in the market without HFT. However, in actual markets, Volume is positively correlated with depth [9], [10]. Yagi et al. [46] suggested that this empirical pattern occurs because the variation in order prices varies dynamically during trading. When the variation in order prices (Est) increases, the number of orders with prices far from the best bid or ask tends to increase, and the amount of executed limit orders in the order book decreases as does the amount of limit orders that are placed outside the measuring range of depth. On the other hand, when the variation in order prices decreases, the number of orders with prices around the best bid or ask tends to increase as does the amount of executed limit orders in the order book and the amount of limit orders that are placed inside the measuring range of depth (see Table VIII).

However, volume may appear to exhibit a positive correlation with depth due to the measuring range of depth. If the range expands, this positive correlation may disappear because depth increases even if the variation in order prices increases.

One of the reasons why volume may exhibit a positive correlation with depth is the variation in the amount of orders sent during a specific period of time. In actual markets, the number of market participants and the amount of orders sent by them change depending on the type of risk assets and the trading time. When the amount of orders changes, the amount of executed orders and the amount of orders that are not executed and remain in the order book will also change. That is to say, the execution rate in the actual market is considered to be almost constant.

Then, by changing the order frequency of normal agents, $\Pr_o$, that is, by changing the amount of orders, we observed how the four market liquidity indicators changed. It is noted that each normal agent places an order with probability $\Pr_o$ and does not with probability $1 - \Pr_o$. The results are shown in Table IX. The findings concur with what has been observed in empirical studies, in that Volume exhibits a positive correlation with depth.

The results suggest that the actual data may include the effect of the amount of orders, and if so, it is necessary to evaluate market liquidity by using the data after removing the influence of the amount of orders. On the other hand, under the condition that the amount of orders is constant, changes in artificial market factors such as the characteristics of the market and the traders affected the execution rate (refer to Tables IV–VIII). In other words, market liquidity is roughly determined by two factors: the amount of orders and execution rate. If the execution rate is constant and the amount of orders increases, volume and depth increase, which agrees with the results of empirical studies. However, if the amount of orders is constant and the execution rate increases, volume increases and depth decreases. In empirical studies, such a result has not been clearly measured, but our research could confirm such a result by analyzing an internal market mechanism, which makes it possible to clarify how the effects of changing parameters spread inside the market, using an artificial market. Therefore, we suggest that the execution rate may be an appropriate market liquidity indicator.

A lower the execution rate corresponds to a higher liquidity. Investors who place market orders can trade in the market with a low execution rate, as orders often remain in the order book without being traded despite the large number of orders so far. Note that a prerequisite is that investors place orders at a rational price. This is because when the execution rate drops due to limit orders whose prices deviate far from the fundamental price, it does not have any positive effect on the current transaction (as shown in Table VIII, depth does not increase although the execution rate decreases). In a future study, we will confirm whether the execution rate can be employed as a market liquidity indicator with actual market data.

Finally, we will explain the mechanism of changes in the four market liquidity indicators and their correlations between the market with HFT and the market without HFT, when the order frequency is changed.

When the order frequency decreases, the number of orders which normal agents place also decreases per the simulation results. Then, as both the amount of orders which match orders in the order book and the amount of orders which do not match them and remain in the order book decrease, both volume and depth also decrease. As the number of orders in the order book decreases, the price with no orders in the order book increases, so once some orders in the order book are executed, the bid-ask spread tends to expand. Therefore, tightness increases when the order frequency decreases. The reason why resiliency increases is because volume, which is the denominator of the P/T ratio, drops sharply, although the volatility does not change much, so that the difference between the highest and lowest prices of a day, which is the numerator of the P/T ratio, is relatively stable.

We will compare each indicator in the market with HFT and the market without HFT. In this comparison, only Tightness is inversely correlated. When an HFT participates, a narrow spread with almost constant width is always provided by the HFT. However, as the number of new orders increases, the HFT's orders are matched to them and it is difficult for them to remain in the order book. As a result, the bid-ask spread widens and tightness increases.

## V. Conclusion

We have investigated changes in four well-known market liquidity indicators and their correlations in a market where an HFT participated and compared the results to those when an HFT did not participate. We found that the HFT's transactions contributed to improve all four market liquidity indicators when the market was stable. Thus, market liquidity is supplied by the HFT's transactions. Furthermore, we also found that the reason why volume correlates with depth in actual markets is not only that the variation in order prices of investors changes dynamically [46] but also because the order frequency of investors changes dynamically. This suggests that market liquidity can be measured not only by the four major market liquidity indicators—volume, tightness, resiliency, and depth—but also by the execution rate.

Future work is as follows. First, we will explore changes in market liquidity indicators and the relationships between them in markets where the fundamental price is unstable, as actual markets are often unstable. Next, we will confirm whether the results obtained using an artificial market are valid when actual data are used. As mentioned in Section I, the validation of the simulation model and the results obtained from it with actual data is an enormous task and would be a separate research project. Therefore, in this article, we focused on investigating how the HFT orders affect market liquidity and the relationship between the market liquidity indicators by using a simulation model that is simple but captures the characteristic of the market and the traders. However, it is necessary to investigate both the simulation model and the results obtained from it with actual data.

DISCLAIMER

The opinions contained herein are solely those of the authors and do not necessarily reflect those of SPARX Asset Management Co., Ltd.

REFERENCES

[1] T. Kurosaki, Y. Kumano, K. Okabe, and T. Nagano, "Liquidity in JGB markets: An evaluation from transaction data," Bank Japan, Tokyo, Japan, Bank Japan Working Paper Series 15-E-2, May 2015. [Online]. Available: https://ideas.repec.org/p/boj/bojwps/wp15e02.html

[2] A. S. Kyle, "Continuous auctions and insider trading," *Econometrica*, vol. 53, pp. 1315–1336, Nov. 1985.

[3] J. Olbrys and M. Mursztyn, "Depth, tightness and resiliency as market liquidity dimensions: Evidence from the polish stock market," *Int. J. Comput. Econ. Econometrics*, vol. 9, no. 4, pp. 308–326, 2019.

[4] R. Y. Goyenko, C. W. Holden, and C. A. Trzcinka, "Do liquidity measures measure liquidity?" *J. Financial Econ.*, vol. 92, no. 2, pp. 153–181, 2009.

[5] K. Y. L. Fong, C. W. Holden, and C. Trzcinka, "What are the best liquidity proxies for global research?" *SSRN Electron. J.*, vol. 21, no. 4, pp. 1355–1401, 2017.

[6] J. Olbrys and M. Mursztyn, "Measuring stock market resiliency with discrete Fourier transform for high frequency data," *Phys. A, Stat. Mech. Appl.*, vol. 513, pp. 248–256, Jan. 2019.

[7] J. Olbrys and M. Mursztyn, "Estimation of intraday stock market resiliency: Short-time Fourier transform approach," *Phys. A, Stat. Mech. Appl.*, vol. 535, Dec. 2019, Art. no. 122413.

[8] Y. Amihud, "Illiquidity and stock returns: Cross-section and time-series effects," *J. Financial Markets*, vol. 5, no. 1, pp. 31–56, Jan. 2002.

[9] J. Muranaga, "Dynamics of market liquidity of Japanese stocks: An analysis of tick-by-tick data of the Tokyo stock exchange," in *Market Liquidity: Research Findings and Selected Policy Implications*, vol. 11. Basel, Switzerland: Bank for International Settlements, 1999, pp. 1–25. [Online]. Available: https://EconPapers.repec.org/RePEc:bis:biscgc:11-13

[10] K. H. Chung, K. A. Kim, and P. Kitsabunnarat, "Liquidity and quote clustering in a market with multiple tick sizes," *J. Financial Res.*, vol. 28, no. 2, pp. 177–195, 2005. [Online]. Available: https://onlinelibrary.wiley.com/doi/abs/10.1111/j.1475-6803.2005.00120.x

[11] L. E. Harris, "Minimum price variations, discrete bid–ask spreads, and quotation sizes," *Rev. Financial Stud.*, vol. 7, no. 1, pp. 149–178, 1994.

[12] H. Bessembinder, "Trade execution costs and market quality after decimalization," *J. Financial Quant. Anal.*, pp. 747–777, 2003.

[13] H.-J. Ahn, J. Cai, K. Chan, and Y. Hamao, "Tick size change and liquidity provision on the Tokyo stock exchange," *J. Jpn. Int. Economies*, vol. 21, no. 2, pp. 173–194, Jun. 2007.

[14] R. Naes and J. A. Skjeltorp, "Order book characteristics and the volume-volatility relation: Empirical evidence from a limit order market," *SSRN Electron. J.*, pp. 408–432, 2006.

[15] Bank for International Settlements. (1999). *Recommendations for Design Liquid Markets*. [Online]. Available: https://www.bis.org/publ/cgfs13.htm

[16] K. Nishizaki, A. Tsuchikawa, and T. Yagi, "Indicators related to liquidity in JGB markets," Bank Japan, Bank Japan Rev., Tokyo, Japan, Series 13-E-3, 2013. [Online]. Available: https://EconPapers.repec.org/RePEc:boj:bojrev:13-e-3

[17] R. Von Wyss, "Measuring and predicting liquidity in the stock market," Ph.D. dissertation, Verlag Nicht Ermittelbar, Univ. St. Gallen, St. Gallen, Switzerland, 2004.

[18] A. Ranaldo, "Intraday market liquidity on the swiss stock exchange," *Financial Markets Portfolio Manage.*, vol. 15, no. 3, pp. 309–327, Sep. 2001.

[19] M. Coppejans, I. Domowitz, and A. Madhavan, "Resiliency in an automated auction," unpublished manuscript, 2004.

[20] D. M. Hmaied, A. Grar, and O. B. Sioud, "Dynamics of market liquidity of tunisian stocks: An analysis of market resiliency," *Electron. Markets*, vol. 16, no. 2, pp. 140–153, May 2006.

[21] D. K. Lo and A. D. Hall, "Resiliency of the limit order book," *J. Econ. Dyn. Control*, vol. 61, pp. 222–244, Dec. 2015.

[22] J. Dong, A. Kempf, and P. K. Yadav, "Resiliency, the neglected dimension of market liquidity: Empirical evidence from the New York stock exchange," Tech. Rep., 2007.

[23] H.-J. Ahn, K.-H. Bae, and K. Chan, "Limit orders, depth, and volatility: Evidence from the stock exchange of Hong Kong," *J. Finance*, vol. 56, no. 2, pp. 767–788, Apr. 2001.

[24] C. M. C. Lee, B. Mucklow, and M. J. Ready, "Spreads, depths, and the impact of earnings information: An intraday analysis," *Rev. Financial Stud.*, vol. 6, no. 2, pp. 345–374, Apr. 1993.

[25] T. Chordia, R. Roll, and A. Subrahmanyam, "Market liquidity and trading activity," *J. Finance*, vol. 56, no. 2, pp. 501–530, Apr. 2001.

[26] A. Sarr and T. Lybek, *Measuring Liquidity in Financial Markets*, vol. 2. Washington, DC, USA: International Monetary Fund, 2002.

[27] R. D. Huang and H. R. Stoll, "Dealer versus auction markets: A paired comparison of execution costs on NASDAQ and the NYSE," *J. Financial Econ.*, vol. 41, no. 3, pp. 313–357, Jul. 1996.

[28] A. Securities et al., "Report 215: 'Australian equity market structure,'" Austral. Secur. Investments Commission, Canberra, ACT, Australia, Tech. Rep. 215, 2010.

[29] G. Hosaka, "Analysis of high-frequency trading at tokyo stock exchange," Japan Exchange Group, JPX Working Papers, Tokyo, Japan, May 2014, Paper 4. [Online]. Available: https://www.jpx.co.jp/english/corporate/research-study/working-paper/index.html

[30] B. Hagströmer and L. Nordén, "The diversity of high-frequency traders," *J. Financial Markets*, vol. 16, no. 4, pp. 741–770, 2013. [Online]. Available: http://www.sciencedirect.com/science/article/pii/S1386418113000256

[31] K. Kanazawa, T. Sueshige, H. Takayasu, and M. Takayasu, "Derivation of the Boltzmann equation for financial Brownian motion: Direct observation of the collective motion of high-frequency traders," *Phys. Rev. Lett.*, vol. 120, no. 13, Mar. 2018, Art. no. 138301.

[32] R. G. Palmer, W. B. Arthur, J. H. Holland, B. LeBaron, and P. Tayler, "Artificial economic life: A simple model of a stockmarket," *Phys. D, Nonlinear Phenomena*, vol. 75, nos. 1–3, pp. 264–274, Aug. 1994.

[33] W. Arthur, J. Holland, B. Lebaron, R. Palmer, and P. Tayler, "Asset pricing under endogenous expectations in an artificial stock market," in *The Economy as an Evolving Complex System II*. Reading, MA, USA: Addison-Wesley, 1997, pp. 15–44.

[34] C. Chiarella, G. Iori, and J. Perelló, "The impact of heterogeneous trading rules on the limit order book and order flows," *J. Econ. Dyn. Control*, vol. 33, no. 3, pp. 525–537, Mar. 2009.

[35] S.-H. Chen, C.-L. Chang, and Y.-R. Du, "Agent-based economic models and econometrics," *Knowl. Eng. Rev.*, vol. 27, no. 2, pp. 187–219, Apr. 2012.

[36] T. Mizuta, K. Izumi, I. Yagi, and S. Yoshimura, "Regulations' effectiveness for market turbulence by large erroneous orders using multi agent simulation," in *Proc. IEEE Conf. Comput. Intell. Financial Eng. Econ. (CIFEr)*, Mar. 2014, pp. 138–143.

[37] P. Ye, S. Wang, and F.-Y. Wang, "A general cognitive architecture for agent-based modeling in artificial societies," *IEEE Trans. Comput. Social Syst.*, vol. 5, no. 1, pp. 176–185, Mar. 2018.

[38] A. Trivedi and S. Rao, "Agent-based modeling of emergency evacuations considering human panic behavior," *IEEE Trans. Comput. Social Syst.*, vol. 5, no. 1, pp. 277–288, Mar. 2018.

[39] R. L. Rizzi, W. L. Kaizer, C. B. Rizzi, G. Galante, and F. C. Coelho, "Modeling direct transmission diseases using parallel bitstring agent-based models," *IEEE Trans. Comput. Social Syst.*, vol. 5, no. 4, pp. 1109–1120, Dec. 2018.

[40] N. Sardana, R. Cohen, J. Zhang, and S. Chen, "A Bayesian multiagent trust model for social networks," *IEEE Trans. Comput. Social Syst.*, vol. 5, no. 4, pp. 995–1008, Dec. 2018.

[41] I. Yagi, T. Mizuta, and K. Izumi, "A study on the effectiveness of short-selling regulation in view of regulation period using artificial markets," *Evol. Institutional Econ. Rev.*, vol. 7, no. 1, pp. 113–132, 2010.

[42] T. Mizuta, K. Izumi, I. Yagi, and S. Yoshimura, "Design of financial market regulations against large price fluctuations using by artificial market simulations," *J. Math. Finance*, vol. 3, no. 2, pp. 15–22, 2013.

[43] A. Nozaki, T. Mizuta, and I. Yagi, "A study on the market impact of the rule for investment diversification at the time of a market crash using a multi-agent simulation," *IEICE Trans. Inf. Syst.*, vol. E100.D, no. 12, pp. 2878–2887, 2017.

[44] I. Yagi, A. Nozaki, and T. Mizuta, "Investigation of the rule for investment diversification at the time of a market crash using an artificial market simulation," *Evol. Institutional Econ. Rev.*, vol. 14, no. 2, pp. 451–465, Dec. 2017. [Online]. Available: https://link.springer.com/article/10.1007%2Fs40844-017-0070-9

[45] L. Ponta, E. Scalas, M. Raberto, and S. Cincotti, "Statistical analysis and agent-based microstructure modeling of high-frequency financial trading," *IEEE J. Sel. Topics Signal Process.*, vol. 6, no. 4, pp. 381–387, Aug. 2012.

[46] I. Yagi, Y. Masuda, and T. Mizuta, "Detection of factors influencing market liquidity using an agent-based simulation," in *Network Theory and Agent-Based Modeling in Economics and Finance*. Singapore: Springer, 2019, pp. 111–131.

[47] G. Iori, "A microsimulation of traders activity in the stock market: The role of heterogeneity, agents' interactions and trade frictions," *J. Econ. Behav. Org.*, vol. 49, no. 2, pp. 269–285, 2002.

[48] C. Chiarella and X.-Z. He, "Heterogeneous beliefs, risk, and learning in a simple asset-pricing model with a market maker," *Macroeconomic Dyn.*, vol. 7, no. 4, pp. 503–536, 2003.

[49] S. Ganesh, N. Vadori, M. Xu, H. Zheng, P. Reddy, and M. Veloso, "Reinforcement learning for market making in a multi-agent dealer market," 2019, *arXiv:1911.05892*. [Online]. Available: http://arxiv.org/abs/1911.05892

[50] B. LeBaron, "Agent-based computational finance," *Handbook Comput. Econ.*, vol. 2, pp. 1187–1233, 2006.

[51] M. Hirano, K. Izumi, H. Matsushima, and H. Sakaji, "Comparing actual and simulated HFT traders' behavior for agent design," *J. Artif. Societies Social Simul.*, vol. 23, no. 3, 2020. [Online]. Available: http://jasss.soc.surrey.ac.uk/23/3/6.html

[52] Y. Kusada, T. Mizuta, S. Hayakawa, and K. Izumi, "Impact of position-based market makers to shares of markets volumes—An artificial market approach," (in Japanese), *Trans. Jpn. Soc. Artif. Intell.*, vol. 30, no. 5, pp. 675–682, 2015.

[53] T. Lux and M. Marchesi, "Scaling and criticality in a stochastic multi-agent model of a financial market," *Nature*, vol. 397, no. 6719, pp. 498–500, 1999.

[54] T. Nakada and K. Takadama, "Analysis on the number of XCS agents in agent-based computational finance," in *Proc. IEEE Conf. Comput. Intell. Financial Eng. Econ. (CIFEr)*, Apr. 2013, pp. 8–13.

[55] Y. Kusada, T. Mizuta, S. Hayakawa, and K. Izumi, "Impacts of position-based market makers on markets' shares of trading volumes—An artificial market approach," in *Proc. Social Modeling Simulations Econophys. Colloq*, 2014, pp. 1–2.

[56] Y. Nakajima and Y. Shiozawa, "Usefulness and feasibility of market maker in a thin market," in *Proc. Int. Conf. Exp. Econ. Sci. New Approaches Solving Real-World Problems*, Dec. 2004, pp. 1000–1003. [Online]. Available: https://www.cc.kyoto-su.ac.jp/project/orc/execo/EES2004/proceedings.html

[57] M. Sewell. (2006). *Characterization of Financial Time Series*. [Online]. Available: http://financ.emartinsewell.com/stylized-facts/

[58] R. Cont, "Empirical properties of asset returns: Stylized facts and statistical issues," *Quant. Finance*, vol. 1, no. 2, pp. 223–236, Feb. 2001.

[59] R. Tsay, *Analysis of Financial Time Series* (Wiley Series in Probability and Statistics), 2nd ed. Hoboken, NJ, USA: Wiley, 2005.

[60] H. Levy, M. Levy, and S. Solomon, *Microscopic Simulation of Financial Markets: From Investor Behavior to Market Phenomena*. Amsterdam, The Netherlands: Elsevier, 2000.

[61] M. Raberto, S. Cincotti, S. M. Focardi, and M. Marchesi, "Agent-based simulation of a financial market," *Phys. A, Stat. Mech. Appl.*, vol. 299, nos. 1–2, pp. 319–327, Oct. 2001.

[62] B. LeBaron, "Agent-based computational finance: Suggested readings and early research," *J. Econ. Dyn. Control*, vol. 24, nos. 5–7, pp. 679–702, Jun. 2000.

[63] C. Chiarella and G. Iori, "A simulation analysis of the microstructure of double auction markets," *Quant. Finance*, vol. 2, no. 5, pp. 346–353, Oct. 2002.

[64] C. H. Hommes, "Heterogeneous agent models in economics and finance," in *Handbook of Computational Economics*, vol. 2. Amsterdam, The Netherlands: North Holland, 2006, pp. 1109–1186.

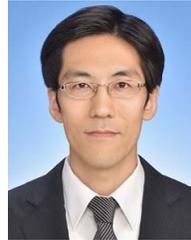

**Isao Yagi** (Member, IEEE) received the Ph.D. degree in information science from the Nara Institute of Science and Technology, Nara, Japan, in 2006.

He was a Research Assistant Professor with the Nara Institute of Science and Technology from 2006 to 2009. He was a Post-Doctoral Fellow with the Tokyo Institute of Technology, Tokyo, Japan, from 2009 to 2011. He was an Associate Professor with the Kanagawa Institute of Technology, Kanagawa, Japan, from 2011 to 2020, where he is currently a Professor. His current research interests include agent-based modeling and social simulation.

**Yuji Masuda** received the B.E. and M.E. degrees from the Kanagawa Institute of Technology, Kanagawa, Japan, in 2018 and 2020, respectively.

He is with Yamato System Development Company Ltd. His research has focused on multiagent simulation, and artificial stock market simulation.

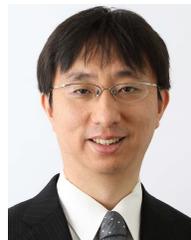

**Takanobu Mizuta** received the B.E. degree from the Meteorological College, Kashiwa, Japan, in 2000, the M.E. degree from the Graduate School of Science, University of Tokyo, Tokyo, Japan, in 2002, and the Ph.D. degree from the School of Engineering, University of Tokyo, in 2014.

Since 2004, he has been with SPARX Asset Management Company Ltd., where he is currently a Fund Manager and a Senior Researcher. Since 2014, he has been a Part-Time Lecturer with the Graduate School of Public Policy, University of Tokyo.